\documentclass[superscriptaddress,prd,aps,twocolumn,floats,nofootinbib,amsfonts,amssymb,preprintnumbers,showpacs]{revtex4}
\usepackage{mathrsfs,graphicx,bm,amsmath}
\begin{document}
\title{\bf Cosmological bounds on the equation of state of dark matter}
\author{Christian M. M{\"{u}}ller}
\affiliation{Institut f{\"{u}}r Theoretische Physik, Philosophenweg
16, 69120 Heidelberg, Germany}
\date{October 26th 2004}
\pacs{98.80-k, 95.35+d}

\newcommand{\be}{\begin{equation}}
  \newcommand{\ee}{\end{equation}}
\newcommand{\ba}{\begin{eqnarray}}
  \newcommand{\ea}{\end{eqnarray}}

\newcommand{\m}{M_{\bar{P}}}
\newcommand{\atoa}{\frac{\dot a}{a}}

\begin{abstract}
In this exploratory study, we investigate the bounds on the equation of state of dark matter.
Modeling dark matter as a fluid component, we
take into account both positive and negative fixed equations of state. Using CMB, supernovae Ia and
 large scale
structure data we find constraints on the equation of state in a modified $\Lambda$CDM cosmology.
 We obtain 
$-1.50  \times 10^{-6} < w_{dm} < 1.13 \times 10^{-6}$ if the dark matter produces no entropy and
 $ -8.78 \times 10^{-3}< w_{dm} < 1.86 \times 10^{-3}$ if the adiabatic sound speed vanishes, both at
3 $\sigma$ confidence level.   
\end{abstract}
\preprint{HD--THEP-04-44}

\maketitle


\section{Introduction}
In this note, we would like to present bounds on the dark matter equation of state 
derived from cosmological observations. 
In other words: given the information from present cosmological observations, how cold is cold dark 
matter? In this context, we do not want to limit ourselves to a non-negative equation
of state since the nature of dark matter is not yet clear.
There are a number of particle dark matter candidates and numerous experiments
attempting to detect dark matter particles ( see \cite{Bertone:2004pz} and references therein). However, 
 dark matter may not be a particle at all. We will try to keep an open mind and 
check if a negative equation of state can be ruled out by cosmological observations.
Previous studies focused on the power spectrum 
properties of warm dark matter \cite{Colombi:1995ze} and mixed dark matter models \cite{Davis:1992ui},
but we would like to emphasize that the present work does not consider the bounds on warm dark matter.
We will consider both positive and negative equations of state $w_{dm}$ and show that the equation of state of dark matter is already strongly
constrained by current observations of the CMB, supernovae Ia and large scale structure.
 We will not concern ourselves 
with bounds from other than cosmological observations. 

This is meant as an exploratory study, not as rigid modeling of dark matter, and we will
therefore work only from a fluid perspective and leave the question open how one could 
obtain a negative equation of state of dark matter. For clarity, we will only allow for a 
constant equation of state $w_{dm}$, even though from a particle perspective it is clear
that $w_{dm}$ varies with time for warm and mixed dark matter models. We have chosen this approach
as there is no simple possibility to give a corresponding particle motivation for $w_{dm}<0$, as specifying
the time evolution of $w_{dm}$ would be model dependent.

In models where dark matter interacts with other components of the universe, such as
in coupled quintessence models \cite{Amendola:1999er}, 
one may obtain a negative equation of state for 
dark matter. It may be possible to obtain $w_{dm} <0$ by other methods as well, but we
are not aware of any such model.

In this work we will
use two simple models for dark matter: one with no entropy production and one with vanishing 
adiabatic sound speed, both with fixed equation of state.

We obtain bounds for a constant equation of state of dark matter of
 $-1.50  \times 10^{-6} < w_{dm} < 1.13 \times 10^{-6}$ if 
there is no entropy production and $ -8.78 \times 10^{-3}< w_{dm} < 1.86 \times 10^{-3}$ if the adiabatic
 sound speed vanishes, both at
3 $\sigma$ confidence level.

For this investigation we will assume that the universe is flat and contains a cosmological constant type dark 
energy component with equation of state $w_{de}=-1$,  dark matter with
a variable equation of state $w_{dm}$, baryons, photons and massless neutrinos. We do not include the tensor part in 
our analysis. In the conclusions, we will address the issue what will change if we relax the flatness 
assumption and if we have an equation of state different from $-1$.  

The plan of this note is as follows:
in Section \ref{sec::pert}, we will introduce the perturbation equations for dark matter with an
arbitrary equation of state (which we will from now on call modified dark matter for brevity). In Section 
\ref{sec::gamma} we introduce a model with vanishing entropy production, in Section \ref{sec::sound}
we give a different model with vanishing adiabatic sound speed. In Section \ref{sec::MCMC} we
present the bounds on $w_{dm}$ obtained from a Markov Chain Monte Carlo simulation, followed 
by our conclusions.  
\section{Perturbation equations}\label{sec::pert}
In the following, we will use the notation of \cite{Kodama:1985bj}. Since this reference may not be readily
available, we will shortly introduce our notation. The full energy momentum tensor for a fluid with
equation of state $w$ may expressed by (we are suppressing
the species index for notational convenience)
\begin{eqnarray}
T_{\;\;0}^0&=& -\bar{\rho} (1+ \delta Q), \\
T^i_{\;\;0}&=& -\bar{\rho} (1+w)\, v\, Q^i,  \\
T^0_{\;\;i}&=& \bar{\rho} \,(1+w)\, (v-B)\, Q_i, \\
T^i_{\;\;j}&=&\bar{p} \,\left[\left(1+\pi_LQ \right) \delta^i_{\,j} + \Pi Q^i_{\,j} \right] \label{eqn::defPi},
\end{eqnarray}
where the $Q(\bm{k},\bm{x})$ are eigenfunctions of the Laplace-Operator, $\nabla^2 Q_{k}(\bm{x})=-k^2 Q_{k}(\bm{x})$ and
in spatially flat universes $Q=\exp(i \bm{k}\bm{x})$.
 The (gauge dependent) variables $\delta, v, B, \pi_L$ and $\Pi$ are defined by these equations. These quantities may then be combined to form the gauge independent quantities $\Delta_g$, 
the density perturbation on hypersurfaces 
of constant curvature perturbation, $V$, the velocity and $\Pi$, the anisotropic stress component. 
We will construct a gauge invariant form of the pressure perturbation $\pi_L$ later. 

The perturbation equations for the dark 
matter component expressed in gauge-invariant variables are then \cite{Kodama:1985bj,Doran:2003xq}\footnote{Please note that w is the equation of state of dark matter and that the species indices are suppressed.}:
\ba 
\dot \Delta_g& +& 3(c_{s}^2-w)\atoa \Delta_g \nonumber \\
&+& k V(1+w)
 + 3\atoa w \Gamma = 0, \label{eq::delta} \\
\label{eq::V}
\dot V&=&\atoa(3 c_{s}^2-1)V+k(\Psi-3c_{s}^2\Phi) \nonumber \\
&+& \frac{c_{s}^2 k}{1+w}\Delta_g+
\frac{w k}{1+w}\left(\Gamma- \frac{2}{3}\Pi\right), \\
a^2&\sum_{\alpha}& (\rho_{\alpha} + p_{\alpha})V_{\alpha} =  2 \m^2 k \left(\atoa \Psi - \dot \Phi\right) 
\label{eq::phidot},
\ea
where the sum is over all present species. 
We will assume that the anisotropic stress 
 vanishes for dark matter, $\Pi_{dm} =0$. 
$\Phi$ and $\Psi$ are the gravitational potentials where
$\Phi = -\Psi -\sum_{\alpha }\Omega_{\alpha}\Pi_{\alpha}$. The sound speed is given by $c_s^2=\dot p/\dot \rho$. 
Note that this is not the 
adiabatic sound speed, which is defined by $c_{ad}^2=\delta p/\delta \rho$. $\Gamma$ is the entropy production
rate and is given by
\be
\Gamma = \pi_L - \frac{c_s^2}{w}\delta.
\ee
This may also be expressed as the difference between ``background'' sound speed and adiabatic sound speed,
\be
w \Gamma = (c_{ad}^2-c_s^2)\delta.
\ee
It will be useful to formulate $\Gamma$ in terms of gauge-invariant variables. From \cite{Kodama:1985bj}
it may be verified that
\be \label{eq::pitilde}
\tilde \pi_L := \pi_L + 3\frac{c_s^2}{w}(1+w)\atoa k^{-1}\sigma_g,
\ee
is the gauge-invariant pressure perturbation. Hence
\be
\Gamma = \tilde \pi_L - \frac{c_s^2}{w}[\Delta_g-3(1+w)\Phi].
\ee
\section{Dark matter with no entropy production}\label{sec::gamma}
\begin{figure}
  \begin{center}
    \includegraphics[scale=0.3]{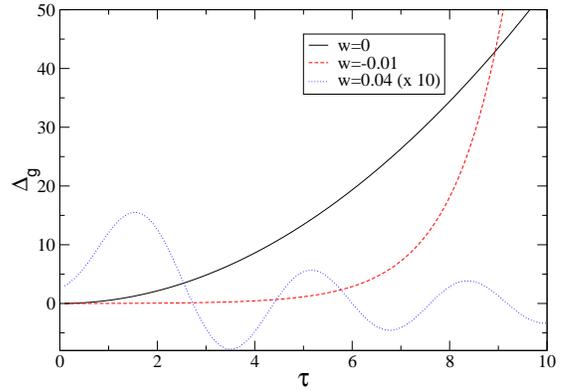}
    \caption{Evolution of the energy density perturbation of dark matter $\Delta_g$ in the sub-horizon regime for
	several equations of state in a universe containing only dark matter with $\Gamma=0$ .
 In the case of $w=0$, the energy density perturbation grows $\propto \tau^2$ (black, straight).
For $0<w<1/3$ (blue, dotted) one obtains decaying oscillations
	while for $w<0$ the density perturbation grows rapidly (red, dashed). }
    \label{fig::analytic}
  \end{center}
\end{figure}

In this section, we will assume that $\Gamma=0$. For a constant equation of 
state of dark matter, the sound speed is given by 
\be
c_s^2=\dot p/\dot \rho=w.
\ee
If we allow for negative equation of state the square of the background sound 
speed may thus become negative. There is of course a question
whether or not this assumption is reasonable, but given the unknown nature
of dark matter we may consider this case and see how well constrained $w$ is. 
Before we discuss the numerical solutions of the perturbation equations, it 
will be helpful to consider the solutions for a universe filled only with 
dark matter in the sub-horizon limit, $k^2\gg \left(\atoa\right)^2$.
 Equations (\ref{eq::delta})-(\ref{eq::phidot}) may be combined to eleminate
$V$:
\ba \label{eq::delta_second_order}
&&\ddot \Delta_g -(3 w-1)\atoa \dot\Delta_g  + w k^2 \Delta_g \nonumber \\ && \hspace{1cm} + k^2(1+w)(\Psi-3w\Phi) =0, \\
&&\dot \Phi \left(\atoa\right) +\left({\frac{5}{2}+
\frac{3}{2}}w\right)\left( \atoa\right)^{2}\Phi \nonumber \\ 
&&\hspace{2.5cm} +\frac{k^2}{3}\Phi=\frac{\Delta_g}{2}\left( \atoa\right)^{2}. \label{eq::phi}
\ea
The background solution for a universe filled only with a modified dark matter component 
yields 
\be
\atoa = \frac{2}{\tau + 3\tau w}.
\ee
For $w=0$ the solution of (\ref{eq::delta_second_order}) and (\ref{eq::phi}) in the 
sub-horizon limit is $\Delta_g=a(\tau)=\tau^2$ and 
$\Phi=\mbox{const.} $, as is well known. For the superhorizon regime, $\Delta_g=\mbox{const.}$
The solutions to these equations in the sub-horizon limit if $w\neq 0$ are plotted in Fig.~\ref{fig::analytic}. 
It can be seen
that for $0<w<1/3$ we obtain decaying oscillations for $\Delta_g$ while for $w=1/3$ one gets oscillations
with constant amplitude. For $w<0$ we find that $\Delta_g$ increases rapidly. This is obviously due to 
the negative sound speed, which leads to a growing gravitational potential. We can thus already see 
that the equation of state will be strongly constrained in the regime $w<0$ since we do not observe this excess 
of power on small scales. Also, the growing gravitational potential would lead to excessive 
lensing on small scales, which
is not observed. At super-horizon scales, $\Delta_g=\mbox{const.}$, regardless
of the equation of state. 

\begin{figure}
  \begin{center}
   \includegraphics[scale=0.35]{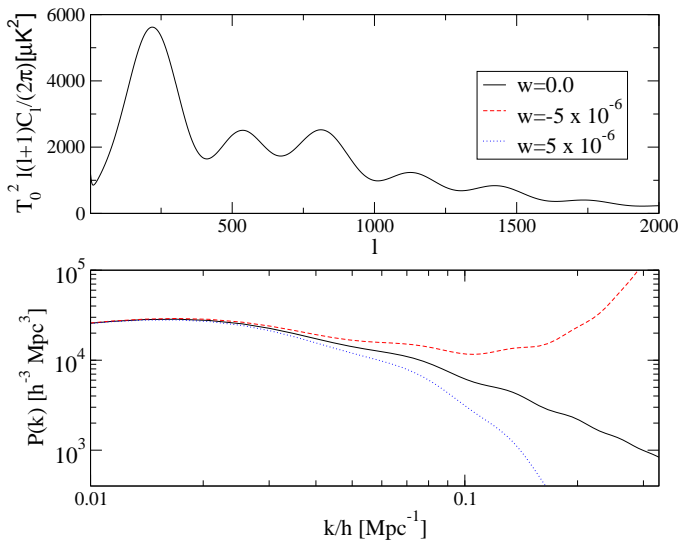}
    \caption{CMB and matter power spectra for a $\Lambda$CDM model with modified dark matter, $\Gamma=0$.
	The power on small scales is much larger than in the $w=0$ case if the equation of 
state is negative (red, dashed). 
	One can see the decaying oscillations for the case of $w>0$ (blue, dotted). 
As expected, there is no difference in the power spectra at very large scales.
The impact on the CMB spectrum is 
not noticable at this level.}
    \label{fig::power_gamma}
  \end{center}
\end{figure}

We have computed the resulting CMB and power spectrum in Fig.~\ref{fig::power_gamma} with the \textsc{Cmbeasy} 
software \cite{Doran:2003sy} (which is based on \textsc{Cmbfast} \cite{Seljak:1996is}), using
 the parameters of a best fit $\Lambda$CDM model:
 $\Omega_b h^2=0.230$, $\Omega_m h^2=0.144$,
$n_s=0.961$, $\tau=0.114$, $h=0.69$.

As one can verify the impact of varying $w$ on the power spectrum is very marked on small
scales. We therefore expect to obtain tight constraints on $w$.
\section{Dark matter with vanishing adiabatic sound speed}\label{sec::sound}
Given the results of the last section and the problematic assumption of negative sound speed, we 
may choose a different approach. In fact, $\Gamma$ measures the ``difference'' between adiabatic sound speed 
$c_{ad}^2=\delta p/\delta \rho$ and ``background'' sound speed $c_s^2=\dot p/\dot \rho$. In the
 previous example, we enforced
$c_{ad}^2=c_s^2$ by the requirement $\Gamma=0$. A different requirement would be that the adiabatic sound
 speed vanishes. 

The first problem that arises is that $c_{ad}^2$ is not gauge-invariant. We must therefore first specify 
what we mean
by vanishing adiabatic sound speed. We may choose a hypersurface such that $c_{ad}^2=0$ has a definite meaning.
Since $c_{ad}^2=0$ implies $\delta p=0$ and $\pi_L=\delta p/p$ this leads to the requirement that $\pi_L=0$ on
a certain set of hypersurfaces. For simplicity, we choose the Newtonian slicing, giving the
shear free hypersurfaces, $\sigma_g=0$. We therefore obtain $\pi_L^{(newt)}=0$, which with the definition
Eq. (\ref{eq::pitilde}) gives
\be
\tilde \pi_L = 0.
\ee
This is true in any gauge because $\tilde \pi_L$ is gauge-invariant. Hence choosing 
\be
\Gamma= 3(1+w)\Phi-\Delta_g,
\ee
we enforce that the adiabatic sound speed vanishes on hypersurfaces of isotropic expansion rate. 
Of course, this choice is by no means preferred over any other choice of hypersurfaces; 
we have chosen this one merely for computational simplicity.

Solving the perturbation equations for a universe filled only with modified dark matter, we obtain
evolution of super-horizon modes if $w \neq 0$. There is no exponential growth for sub-horizon modes
if $w<0$. We may conclude that this model is well-behaved compared to the $\Gamma=0$ case. 

\begin{figure}
  \begin{center}
    \includegraphics[angle=-90,scale=0.35]{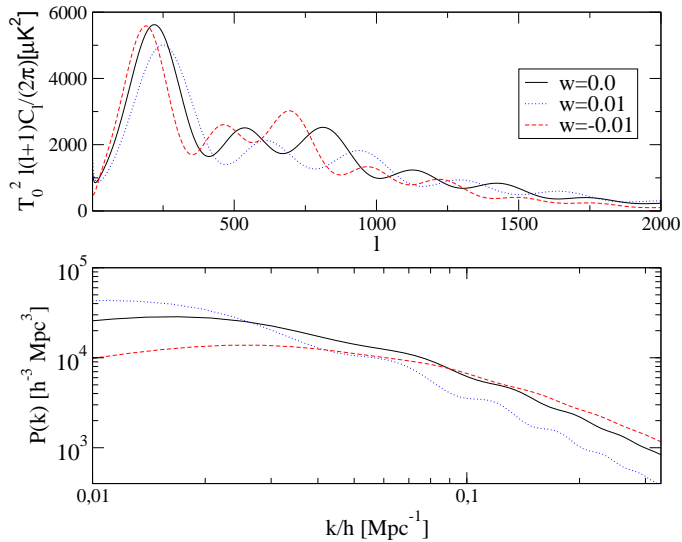}
    \caption{CMB and power spectrum for a $\Lambda$CDM model with modified dark 
matter, $c_{ad}^2=0$. For a positive equation of state (blue, dotted) the peak
 positions of 
the CMB are shifted to smaller scales and have less power than in the $w=0$ case 
(black, straight). For a negative $w$ (red, dashed), the peak positions are 
shifted to smaller $l$. Note also the different peak ratios in these cases. The 
shapes of the matter power spectrum are different in each case, but in 
contrast to  $\Gamma=0$, there is no dramatic difference at small 
scales with respect to the $w=0$ spectrum. The power spectra at very large
scales are different with respect to each other, indicating evolution of super-horizon sized modes. }
    \label{fig::power_sound}
  \end{center}
\end{figure}

We have plotted the power and CMB spectra in Fig.~\ref{fig::power_sound}( model 
parameters as for the $\Gamma=0$ case).
 As expected, the modification has a huge impact on the growth behaviour of fluctuations. 
\section{Bounds on the equation of state of dark matter}\label{sec::MCMC}
In order to quantify the bounds on the equation of state of dark matter, we ran 
a Markov Chain Monte Carlo (MCMC) simulation for $\Gamma=0$ and $c_{ad}^2=0$ with the \textsc{AnalyzeThis!}
package \cite{Doran:2003ua} using WMAP TT and TE spectra \cite{Hinshaw:2003ex,Kogut:2003et},
 VSA \cite{Dickinson:2004yr}, CBI \cite{Readhead:2004gy} and ACBAR \cite{Kuo:2002ua} data up to $l=2000$ as well as 
the SDSS power spectrum \cite{Tegmark:2003uf} (all points with $k/h < 0.15 \mbox{ Mpc}^{-1}$) and the SNe
 Ia data of Riess et al.~\cite{Riess:2004nr}.
Each run contained $\sim 50,000 $ points after burn-in removal. The model used was a $\Lambda$CDM cosmology
 with modified dark matter and  parameter priors as shown in Table \ref{tab::priors}.
\begin{table}[h] \caption{\label{tab::priors}Flat priors used for the MCMC simulations}
\begin{ruledtabular}
  \begin{tabular}{ccc}
           parameter    & min & max  \\ \hline           
  $\Omega_b h^2$   & 0.016& 0.03   \\
  $\Omega_m h^2$   &  0.05 &  0.3 \\ 
  $h$   & 0.60  & 0.85      \\ 
  $n_s$        & 0.8  & 1.2 \\ 
  $\tau$          & 0  & 0.9  \\
 $w_{dm}$ for $\Gamma=0$ & $- 4 \times 10^{-6}$& $5\times 10^{-6}$ \\
 $w_{dm}$ for $c_{ad}^2=0$& $ -0.02$ &0.02  \\	 
 \end{tabular}
\end{ruledtabular}
\end{table}

\begin{figure*}
  \begin{center}
    \resizebox{180mm}{!}{\includegraphics{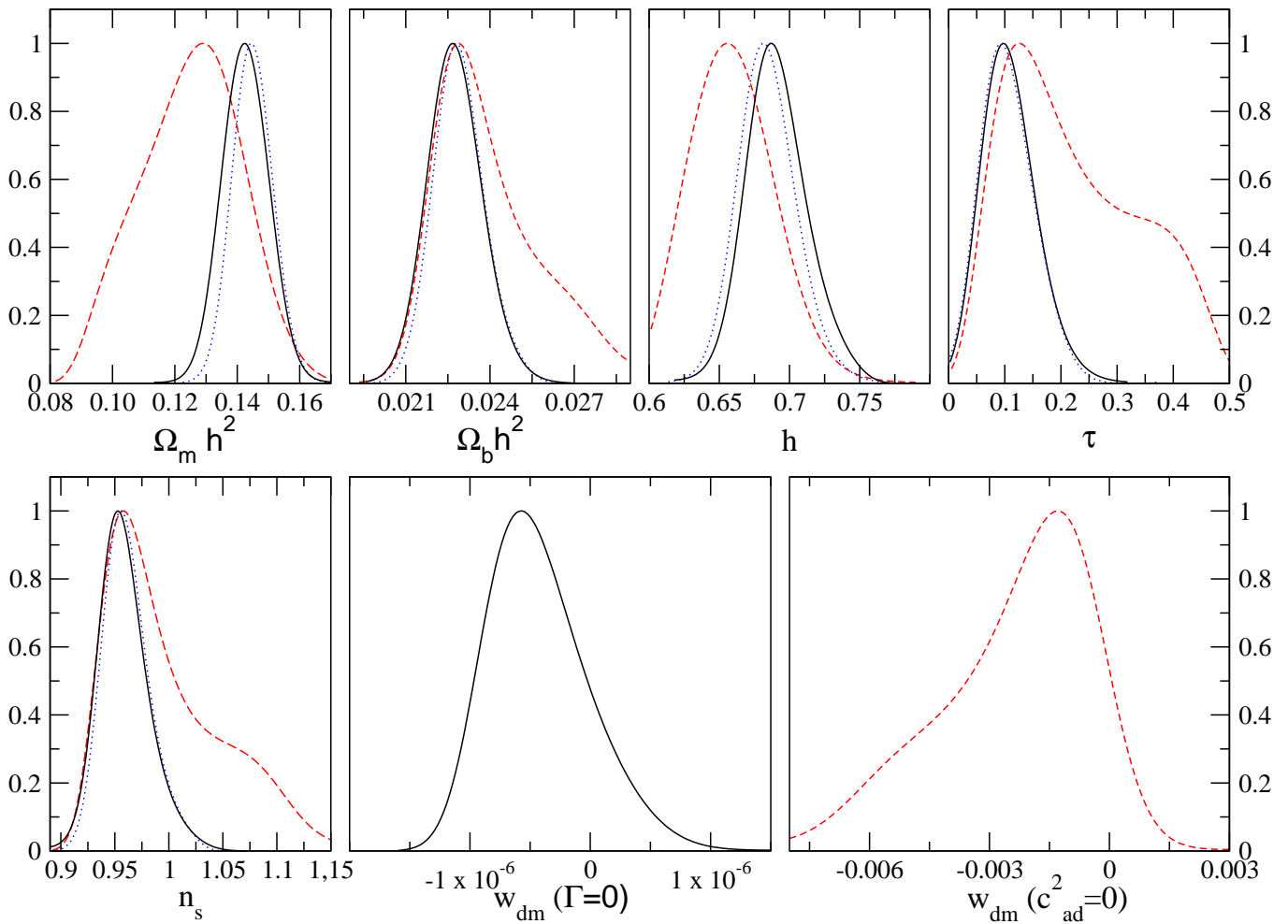}}
    \caption{One dimensional marginalized likelihoods of the MCMC simulation for the
model parameters, using CMB, SNe Ia and LSS data. The results are for the $\Gamma=0$ (black, straight) and 
 the $c_{ad}^2=0$ (red, dashed) model. We have plotted the results for a pure $\Lambda$CDM model 
 with $w=0$ for comparison (blue, dotted). 
Note the large difference in constraints on $w$ for the two models. Also, the $\Gamma=0$ model
has nearly the same parameter distributions as a pure $\Lambda$CDM model.}
    \label{fig::marginalized}
  \end{center}
\end{figure*}

 The resulting one-dimensional marginalized likelihoods are displayed in 
Figure \ref{fig::marginalized}.
 The confidence intervals for the equation of state are displayed in Table \ref{tab::confidence}.


\begin{table}[h] \caption{\label{tab::confidence}Confidence intervals}
\begin{ruledtabular}
  \begin{tabular}{cccc}
             & 68.3 \% & 95.4 \% & 99.7\%  \\ \hline           
 $w_{dm}$ for $\Gamma=0$ ($\times 10^{-6}$)  & $_{-0.929}^{-0.114}$ &
 $_{-1.19}^{+0.462}$ & $_{-1.50}^{+1.133}$ \vspace{0.3cm} \\
 $w_{dm}$ for $c_{ad}^2=0$ ($\times 10^{-3}$)
	& $^{+0.098}_{-3.59}$ & $^{+0.890}_{-6.58}$  & $^{+1.86}_{-8.78}$  \\	 
 \end{tabular}
\end{ruledtabular}
\end{table}
The equation of state is quite strongly constrained if $\Gamma=0$, at a level of $10^{-6}$. What is somewhat 
surprising is that the likelihood is centered not on $w=0$ but on a slightly negative equation of state. 
This may be traced to the fact that the SDSS data set we used does not encompass very small scales and therefore 
the observations are blind to the very strong increase in power at very small scales if $w <0$. The constraints
for the $c_{ad}^2=0$ case are less stringent, in line with our expectations. Here, $w$ is only constrained at a 
level of $10^{-3}$.

\section{Conclusion}\label{sec::conclusions}
Since this is an exploratory study we have not attempted to formulate a model
with realistic variation in the equation of state but chose to show the main effects and bounds
derived from current observations using a constant equation of state.
For the $\Gamma=0$ model, it is clear that the main constraint comes from the 
matter power spectrum. If we would include measurements at smaller scales, the constraints
on the equation of state would be even more restrictive. We may conclude that the $\Gamma=0$ 
model is unlikely to be realistic. 
The situation is not so clear for the $c_{ad}^2=0$ model. More accurate measurements of the CMB, especially
in the large multipole region, should give tighter constraints. Based on the data we cannot conclude that
this model is ruled out. It would be necessary for formulate a specific model before more progress
can be made concerning the question of a possible negative equation of state of dark matter.

How strongly dependent are these results on our assumption of flatness and dark energy equation of state $w_{de}=-1$? From Fig. \ref{fig::power_gamma} we can readily see that the main constraint on the $\Gamma=0$ model comes
from the large scale structure data; since the CMB spectrum does not change much in the allowed parameter 
range, we may conclude that relaxing the flatness assumption would not make much difference on the constraints.
The same is true for an equation of state different from $-1$. From previous studies it is known that for $w_{de}>-1$,
 structure
growth is suppressed at small scales \cite{Caldwell:2003vp,Doran:2001rw,Ferreira:1997hj}.
 But for the $\Gamma=0$ model, this can only make a small difference, since the growth suppression 
cannot  ameliorate the strong deviation from the LSS measurements at small scales as is readily apparent in Fig. 
\ref{fig::power_gamma}.

The situation for the  $c_{ad}^2=0$ case is different. Here, relaxing the flatness assumption and allowing for open or
closed universes  will lead to a 
weaker constraint on $w_{dm}$. The first peak position is sensitive to the geometry of the universe, but increasing 
or decreasing $w_{dm}$ can in principle shift this peak to be in agreement with the WMAP data, as may be 
seen in Fig. \ref{fig::power_sound}. We therefore 
expect also that the constraint on the total energy $\Omega_{tot}$ will be weaker than in  the standard $\Lambda$CDM case. Concerning the possibility that the equation of state of dark energy $w_{de}>-1$ we may say that here, too, the constraints on the dark energy equation of state
and $w_{dm}$ will be less stringent. As mentioned above, the main impact of $w_{de}>-1$ is through a suppression of 
structure growth at small scales, but this may be compensated by decreasing $w_{dm}$ (see Fig. \ref{fig::power_sound}).
It is therefore apparent that allowing for $w_{de} \neq -1$ will result in weaker constraints on $w_{dm}$.

Relaxing the flatness and $w_{de}=-1$ will therefore have a negligible impact for the $\Gamma=0$ model but 
may lead to a significant relaxation of constraints for the $c_{ad}^2=0$ model. 


\begin{acknowledgements}
The author would like to thank C.~Wetterich for suggesting this investigation
 and G.~Sch\"afer and Michael Doran for helpful discussions. C.~M.~M\"{u}ller is supported by 
GRK Grant No.~ 216/3-02. 
\end{acknowledgements}



\bibliographystyle{unsrt}

\end{document}